\documentclass[pra,aps,showpacs,twocolumn,superscriptaddress,preprintnumbers]{revtex4}
\usepackage{graphicx}
\usepackage{dcolumn,epsfig}
\usepackage{amsmath}
\usepackage{hyperref}
\usepackage{color}

\begin{document}
\title{Optical cavities as amplitude filters for squeezed fields}
\author{Thomas Corbitt}
\affiliation{LIGO Laboratory, Massachusetts Institute of
Technology, Cambridge, MA 02139}

\author{Nergis Mavalvala}
\affiliation{LIGO Laboratory, Massachusetts Institute of
Technology, Cambridge, MA 02139}

\author{Stan Whitcomb}
\affiliation{LIGO Laboratory, California Institute of Technology,
Pasadena, CA 02139}

\begin{abstract}
\noindent We explore the use of Fabry-P\'erot cavities as
high-pass filters for squeezed light, and show that they can
increase the sensitivity of interferometric gravitational-wave
detectors without the need for long (kilometer scale) filter
cavities. We derive the parameters for the filters, and analyze
the performance of several possible cavity configurations in the
context of a future gravitational-wave interferometer with
squeezed light (vacuum) injected into the output port.
\end{abstract}

\definecolor{clr}{rgb}{0.6,0,1}
\preprint{\huge \color{clr}{LIGO-P030068-01-R}}

\maketitle

\section{Introduction}
\label{sect:intro} Gravitational-wave interferometers of the type
used in the Laser Interferometer Gravitational-wave Observatory
(LIGO) \cite{LIGO} are typically variants of a Michelson
interferometer with Fabry-P\'erot cavities in the arms. In
addition to the arm cavities, the optical configuration for
interferometers may include a partially reflecting mirror between
the laser source and the input port of the beamsplitter -- a
technique known as power-recycling, and/or a partially
transmitting mirror between the antisymmetric or output port of
the beamsplitter and photodetector -- a technique refered to
signal tuning. The present-day interferometric gravitational-wave
detectors are limited by shot noise at high frequencies ($\gtrsim
100$~Hz); in the next generation of detectors, as other noise
sources are suppressed and the input laser power is increased, the
sensitivity of the detectors is expected to be limited by quantum
optical noise due to fluctuations of the vacuum fields at almost
all frequencies.

Two sources contribute to the limiting optical noise: {\it shot
noise}, which arises from uncertainty due to quantum mechanical
fluctuations in the number of photons at the interferometer
output; and {\it radiation-pressure noise}, which arises from
mirror displacement induced by radiation-pressure fluctuations.
Both sources can be attributed to quantum fluctuations of the
vacuum electromagnetic fields that enter the open ports of the
interferometer \cite{caves1,caves2}. Moreover, they are associated
with orthogonal quadratures of the vacuum: shot noise comes from
the phase quadrature and radiation-pressure noise comes from the
amplitude quadrature of the input vacuum field. The spectrum of
shot noise is independent of frequency, while that of
radiation-pressure noise falls off with increasing frequency,
which causes the two noise sources to dominate in different
frequency bands. An exact balancing of the shot noise and
radiation-pressure noise, in the absence of correlations between
them, leads to the standard quantum limit (SQL).

Injection of squeezed light (vacuum) into the unused output port
of an interferometer was initially proposed as a means of reducing
the shot noise, assuming limited laser power \cite{caves2}, but
the fixed squeeze angle proposed did not breach the SQL. The {\it
squeeze angle} describes the linear combination of input
quadratures in which the fluctuations are reduced; it governs the
relative distribution of quantum noise in each quadrature -- and
hence the quantum noise at all frequencies. Subsequently it was
realized that injection of squeezed vacuum with the proper squeeze
angle would beat the SQL over a narrow bandwidth \cite{unruh}.
Since the radiation-pressure noise dominates at low frequencies
and shot noise at higher frequencies, squeezing with a single
squeeze angle leads to improvement at some frequencies, but
degrades the performance at other frequencies. To achieve
broadband noise reduction using squeezed light, it is necessary to
produce squeezing with a frequency-dependent squeeze angle. Kimble
et al.~\cite{KLMTV01} recognized that squeezed vacuum reflected
from appropriately detuned filter cavities could match the
required squeeze angle over a broad range of frequencies and give
broadband performance below the SQL. Squeezing in signal-tuned
interferometers was analyzed in Refs. \cite{harms}, \cite{spie03}
and \cite{BC5}. Frequency-dependent squeezing using optical
cavities offers excellent performance, but it is likely to be
difficult and costly to implement because it requires long
(kilometer scale) filter cavities to reduce losses so that the
squeezing is not destroyed in the process~\cite{KLMTV01,harms}.

Here we propose using an alternative type of filter cavity that,
instead of giving a frequency-dependent squeeze angle, act as
high-pass filters for the squeeze amplitude. In using this design,
we reduce the harmful effects of squeezing with a constant squeeze
angle and the need for a low-loss cavity, while retaining the
benefits of squeezing at high frequencies. The premise of our
filter is that at high frequencies the input beam will be entirely
reflected by the filter cavity and the squeezing will be
preserved, while at low frequencies, it will be entirely
transmitted and the cavity losses will cause ordinary vacuum to
replace the (anti-) squeezed vacuum noise. The cavities have a
visibility of nearly unity. In general, we attempt to choose a
transition frequency above which the squeezing is preserved and
below which we destroy the anti-squeezing. For any realistic
cavity, however, there will be a transition region in which the
squeezing is also partially destroyed. The use of multiple
cavities allows manipulation, and hence optimization, of this
transition region.

In this paper we evaluate the performance of these filter cavities
and propose configurations that reduce the extent of the
transition region. We initially consider the use of a single
filter cavity, [Fig.~\ref{fig:filt}~(b)]; in
Section~\ref{sect:multiplefilts}, we consider the extension to
multiple cavities. In each case the filter cavities are placed
between the squeeze source and unused output port of the beam
splitter, as shown in Fig.~\ref{fig:filt}~(a). We evaluate the
performance using three astrophysical criteria {\it
simultaneously}: (i) the signal-to-noise ratio (SNR) for detecting
a stochastic background of gravitational-waves, (ii) the
signal-to-noise ratio (SNR) for inspiraling neutron star binaries,
and (iii) the strain sensitivity at higher frequencies, where
pulsars are expected to be detectable.

\section{Filter description}
\label{sect:param} 
\begin{figure}[t]
\includegraphics[width=8.7cm]{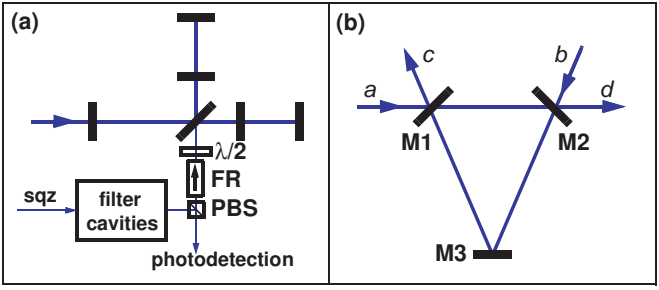}
\caption[Figure] {\label{fig:filt} Panel (a) displays the
placement of the filter in a squeezed-input interferometer. Panel
(b) shows the single filter configuration, where $a$ is the beam
to be filtered, $b$ is unsqueezed vacuum, $c$ is the filtered beam
and $d$ is the transmitted beam.}
\end{figure}

We consider a triangular cavity with three
mirrors, as shown in Fig.~\ref{fig:filt}~(b), where $R_{i}$,
$T_{i}$ and $A_{i}$ are the power reflectivity, transmission and
loss, respectively, of each mirror such that

\begin{equation}
R_{i}+T_{i}+A_{i} = 1 {\rm , \hspace{0.2cm} with \hspace{0.2cm}}
i=1, 2, 3.
\end{equation}

\noindent The field incident on the the cavity comprises a carrier
at frequency, $\omega_0$, and sidebands at frequencies, $\omega_0
\pm \Omega$. When the cavity is resonant with the carrier
frequency, $\omega_0$, the roundtrip length of the cavity, $l$, is
an integer number of carrier half-wavelengths and the
$\omega_{0}+\Omega$ component of the incident field will
experience a frequency-dependent phase shift, $\Phi = \frac{\Omega
l}{c}$ for a single round trip of the cavity. Cavity amplitude
reflection and transmission coefficients are then given by

\begin{eqnarray}
\rho\left(\Omega\right) &= \displaystyle \frac{c}{a} &= \sqrt{R_1}
- \frac{T_1\sqrt{R_2 \, R_3}
\,{\rm e}^{2i\Phi}}{1-\sqrt{R_1 \, R_2 \, R_3}\, {\rm e}^{2i\Phi}},\\
\tau\left(\Omega\right) &= \displaystyle \frac{d}{a} &=
\frac{\sqrt{T_1 \, T_2}\, {\rm e}^{i\Phi}}{1-\sqrt{R_1 \, R_2 \,
R_3}\, {\rm e}^{2i\Phi}}.
\end{eqnarray}

\noindent To make the cavity act as a high-pass filter for the
reflected light at frequency $\Omega$ (relative to the carrier
frequency, $\omega_0$), a cavity with no reflected light at
$\Omega = 0$ is desired, so we require that $\rho(0) = 0$. We
constrain the values of $R_1$ and $R_2 \, R_3$ at a fixed value of
$A_{1}$ by choosing a value for the half-linewidth of the cavity

\begin{eqnarray}
\gamma_f &\equiv& \frac{c}{4l}(1-R_1 \, R_2 \, R_3),\\
R_1 &=& \sqrt{1-\frac{4l\gamma_f}{c}}\left(1-A_{1}\right),\\
R_2 \, R_3 &=& \displaystyle
\frac{\sqrt{1-\frac{4l\gamma_f}{c}}}{\left(1-A_{1}\right)},
\end{eqnarray}
resulting in a reflected beam of the form

\begin{equation}
\rho\left(\Omega\right) = \frac{\sqrt{R}(1- {\rm
e}^{2i\Phi})}{1-\sqrt{1-\frac{4l\gamma_f}{c}}{\rm e}^{2i\Phi}}
\approx \frac{\Omega}{i \gamma_f + \Omega}.
\label{rhoeq}
\end{equation}

\noindent Eqn.~(\ref{rhoeq}) shows that the performance of the
cavity depends only on its linewidth. $R_2 \, R_3$ must be less
than $1-A_{2}-A_{3} \approx 1-2A_{1}$, which requires $A_{1} <
\frac{2l\gamma_f}{3c}$. The finesse of the cavity is $\mathcal{F}
= \frac{\pi c}{2l\gamma}$.

\section{Filtered squeezed states}
Since the cavity is not detuned from resonance and there is no
rotation of the quadratures, it is straightforward to extend this
result to the Caves-Schumaker two-photon formalism
\cite{CavesSchumaker,SchumakerCaves}. Refering to
Fig.~\ref{fig:filt}~(b), $a$ and $b$ are the (complex) amplitudes
of fields at sideband frequency, $\Omega$, incident on mirror M1
and M2, respectively. The field reflected from the cavity, $c$,
has the form

\begin{equation}
c_{i} = \rho a_{i} + \tau b_{i} +
\sqrt{1-|\rho|^{2}-|\tau|^{2}}\,v_{i}\,,
\end{equation}

\noindent where $i = 1,\,2$ and the $v_i$ are the quadrature field
amplitudes of the unsqueezed vacuum that leaks in due to the
losses in the cavity. In the case where the light incident on M2
is also unsqueezed vacuum~\footnote{Since the vacuum fluctuations
entering the cavity due to losses and those due to finite
transmissivity are uncorrelated, we add them in quadrature.}, the
reflected field takes the form

\begin{equation}
c_{i} = \rho a_{i} + \sqrt{1-|\rho|^{2}}\,v_{i}. \label{eqn:c}
\end{equation}

\noindent Now suppose squeezed vacuum is incident on the cavity,
i.e. $a$ is squeezed. Since $\rho(\Omega)$ has the response of a
high pass filter, we see from Eqn.~(\ref{eqn:c}) that at low
frequencies where $\rho(\Omega < \gamma_f) \sim 0$, the second
term dominates and the filter output field, $c$, is in an ordinary
vacuum state given by $v$, while at high frequencies where
$(1-|\rho|^2) \sim 0$, the output field contains primarily the
squeezed input vacuum, $a$.

The reflected beam is, in general, not a pure squeezed
state~\footnote{A ``pure'' squeezed state refers to the case where
the variance of the noise in one quadrature increases by exactly
the same amount as that in the orthogonal quadrature is reduced,
i.e. the area of the noise ellipse is unity. When excess noise is
present in one quadrature, this condition is not satisfied.}. Two
parameters characterize the effects of the cavity on the squeezed
state: the attenuation factor, $\alpha$, and the vacuum leakage,
$\beta$. The attenuation factor measures the effect on the
anti-squeezed quadrature, while the vacuum leakage measures the
effect on the squeezed quadrature by measuring the vacuum noise
that enters the beam. Defining $x=\Omega/\gamma_f$, we find
\begin{eqnarray}
\alpha =& |\rho|^{2} &\approx \frac{x^{2}}{1+x^{2}},\\
\label{alphaeq}
\beta =& 1-|\rho|^{2} &\approx \frac{1}{1+x^{2}},\\
\label{betaeq} \alpha + \beta =& 1.
\end{eqnarray}

\begin{figure}[t]
\begin{center}
\includegraphics[width=8.5cm]{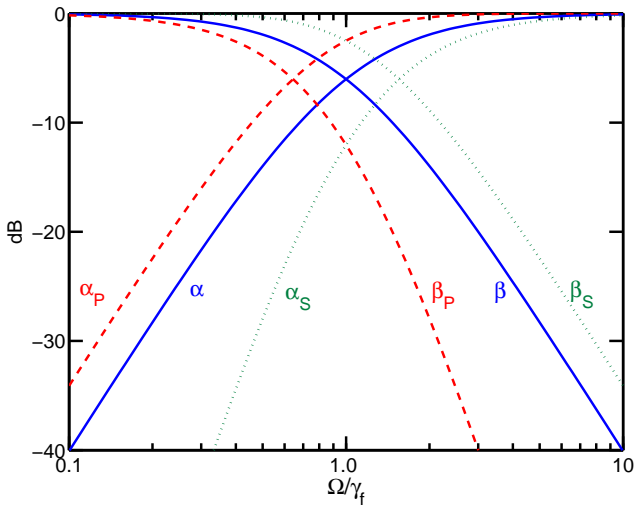}
\caption[Figure] {\label{fig:alphabeta}A plot of the attenuation,
$\alpha$, and maximum possible squeezing, $\beta$ for a single
(blue, solid), series (red, dashed) and parallel (green, dotted)
filter. The subscripts $S$ and $P$ refer to series and parallel
filters, respectively; these filter configurations are described
in Section~\ref{sect:multiplefilts}. The series filter increases
the effective corner frequency and the attenuation factor, while
the parallel filter decreases the effective corner frequency but
increases the maximum squeezing.}
\end{center}
\end{figure}

\noindent We define the corner frequency, $\xi = x_{1/2}
\gamma_f$, of the filter to be the frequency at which $\alpha =
\beta = \frac{1}{2}$, which gives $x_{1/2} \approx 1$. The
parameters $\alpha$ and $\beta$ as a function of (normalized)
frequency are plotted as the solid curves in
Fig.~\ref{fig:alphabeta}. Discussion of multiple filters, also
shown in Fig.~\ref{fig:alphabeta}, is deferred to Section
\ref{sect:multiplefilts}.

\section{Application to gravitational-wave interferometers}
\subsection{Conventional interferometer}
\label{sect:conventional} It is instructive to study the
performance of the amplitude filter cavity with a conventional GW
interferometer \footnote{"Conventional" refers to a power-recycled
Michelson interferometer with Fabry-P\'erot cavities in the arms,
after Ref.~\cite{KLMTV01}.} operated at the optimum power required
to reach the standard quantum limit at $\Omega =
\gamma$~\cite{KLMTV01}. Here $\gamma$ is the linewidth of the arm
cavities of the interferometer, distinct from $\gamma_f$, which is
the linewidth of the filter cavity. Though $\gamma$ may be
adjusted to optimize the detector performance under certain
circumstances, we restrict ourselves to $\gamma = 2 \pi \times 100
{\rm ~Hz}$.

A phase-squeezed vacuum beam is filtered by the cavity and then
injected into the otherwise unused output port of the
interferometer, as shown in Fig.~\ref{fig:filt}~(a). To maintain
the same input/output relations as developed in
Ref.~\cite{KLMTV01}, the interferometer input (cavity output) beam
is given in the form~\footnote{A unitary transformation was
applied to the operator equation $a_{i}^{\prime} = \sqrt{\alpha}\,
a_{i} + \sqrt{\beta}\,v_{i}$ acting on a squeezed state $|{\rm
sqz}\rangle$, such that $a_{1,2} \rightarrow {\rm e}^{\pm R} \,
a_{1,2}$ and $|{\rm sqz}\rangle \rightarrow |0\rangle$.}

\begin{eqnarray}
a_{1}^{\prime} &=& \sqrt{\alpha}\, {\rm e}^{R}a_{1} + \sqrt{\beta}\,v_{1}\\
a_{2}^{\prime} &=& \sqrt{\alpha}\, {\rm e}^{-R}a_{2} +
\sqrt{\beta}\,v_{2}.
\end{eqnarray}

\noindent For an interferometer with arm lengths, $L$, mirror
masses, $m$, and injection squeeze factor, $R$, this leads to the
noise spectral density

\begin{equation}
S_{h} = \frac{h_{SQL}^{2}}{2\kappa}\left [\alpha \left ({\rm
e}^{-2R}+\kappa^{2} {\rm e}^{2R}\right )+\beta \left
(1+\kappa^{2}\right )\right ], \label{strain}
\end{equation}

\noindent where
\begin{equation}
h_{SQL}\equiv \displaystyle \sqrt{\frac{8 \hbar}{m \, \Omega^2 \,
L^2}}
\end{equation}

\noindent is the noise spectral density of the dimensionless
gravitational-wave strain at the standard quantum limit for an
interferometer with uncorrelated shot noise and radiation-pressure
noise, and

\begin{equation}
\kappa = \displaystyle \frac{2 \left(I_0/I_{SQL} \right
)\,\gamma^4}{\Omega^2 \left(\gamma^2 + \Omega^2 \right)}
\end{equation}

\noindent is the effective coupling constant that relates the
output signal to the motion of the interferometer mirrors
\cite{KLMTV01}.

\begin{figure}[t]
\begin{center}
\includegraphics[width=8cm]{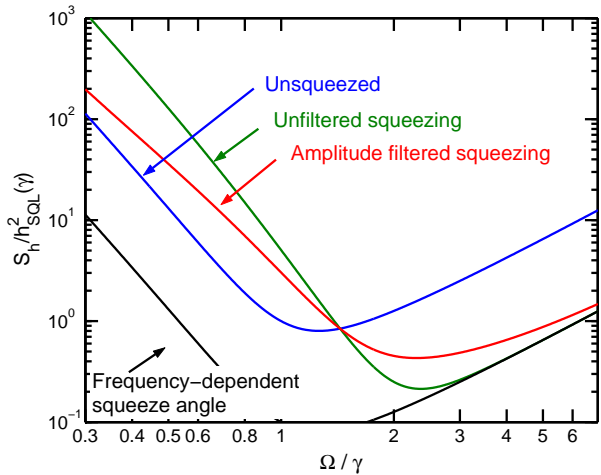}
\end{center}
\caption[Figure] {\label{fig:SNR} The noise spectral density,
normalized by $h_{SQL}$, for a {\it conventional} interferometer
with (i) no squeezed input (``Unsqueezed''); (ii) squeezed vacuum
injected (``Unfiltered Squeezin''); (iii) squeezed input filtered
by a filter cavity (``Amplitude filtered squeezing''); and (iv)
frequency-dependent squeeze angle (``Frequency-dependent squeeze
angle'').}
\end{figure}

In Fig.~\ref{fig:SNR}, we plot the noise spectral densities for a
conventional interferometer with squeezed input parameter ${\rm
e}^{-2\,R} = 0.1$, using different filter cavity configurations.
The unfiltered squeezed input gives significantly higher noise at
$\Omega \lesssim \gamma$. When the squeezed input is filtered by a
single filter, there is a frequency band $\Omega/\gamma \gtrsim
1.5$ in which the sensitivity is worse than the the unfiltered
squeezed case, and a corresponding range $\Omega/\gamma \lesssim
1$ in which the sensitivity is worse than the unsqueezed case. We
refer to this band around $\Omega/\gamma \sim 1$ as the
``transition region''; it is a consequence of the frequency
response of the filter. In Section \ref{sect:multiplefilts} we
examine some methods to reduce the frequency extent of this
transition region such that the low-frequency noise of the
squeezed input interferometer approaches that of an interferometer
with no squeezing, while preserving the noise reduction at high
frequencies.

It is also possible to define a critical frequency above which
squeezing is desirable, and below which it is
deleterious~\footnote{The critical frequency was first pointed out
by Yanbei Chen.}. Inserting $\alpha + \beta = 1$ into
eqn.~(\ref{strain}) for $S_h$, we get

\begin{equation}
S_{h} = \frac{h_{SQL}^{2}}{2\kappa} \left \{\alpha \left [\left
({\rm e}^{2R} - 1 \right) \,\kappa^{2} - \left(1 - {\rm
e}^{-2R}\right )\right] + \left (1 + \kappa^2 \right) \right \}.
\label{strain02}
\end{equation}

\noindent The coefficient of the term in $\alpha$ switches sign at
the critical frequency, $\Omega/\gamma = 1.44$, which corresponds
to the frequency at which the curves in Fig.~\ref{fig:SNR} cross.
Since $\beta = 0$ and $\alpha = 1$ when no filtering is applied,
we always have $\alpha < 1$ in the filtered case. More generally
then, the noise for the unfiltered case is better at frequencies
higher than the critical frequency (where $\kappa$ is small), and
worse at frequencies below the critical frequency. Moreover, at
the critical frequency, the coefficient of $\alpha$ in
eqn.~(\ref{strain02}) is zero and the value of $\alpha$ does not
matter. This also explains why the curves for a variety of filter
configurations in Fig.~\ref{fig:SNR_adligo} all cross at a single
frequency. It will become evident in
Section~\ref{sect:performance}, however, that in optimizing the
filter performance using astrophysical criteria, the critical
frequency does not play a significant role.

\subsection{Signal-recycled interferometer}

The amplitude filter cavities can also be used in conjunction with
a squeezed-input signal-recycled interferometer. The introduction
of signal recycling has the effect of mixing the quadratures in
the input/output relations of the interferometer. This allows the
shot noise and radiation-pressure noise to become correlated,
typically producing two resonances~\cite{BC1}: the lower frequency
dip is due to an optical-mechanical coupling and the higher
frequency one is a purely optical resonance due to the storage
time of the gravitational wave signal in the interferometer. The
mixing of the quadratures results in the optimal squeeze angle
having a strong frequency dependence compared to the conventional
interferometer~\cite{harms}.

\begin{figure}[t]
\begin{center}
\includegraphics[width=8cm]{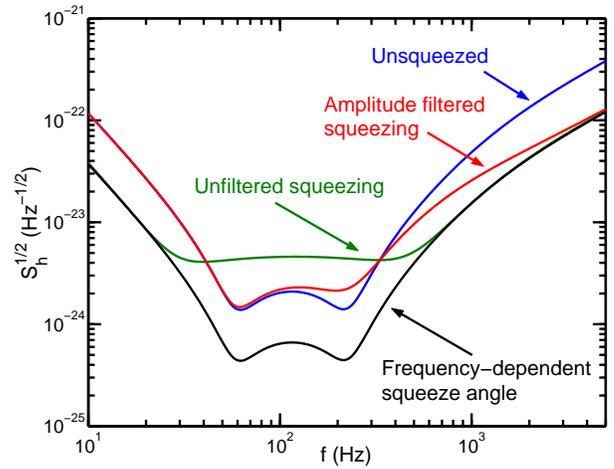}
\end{center}
\caption[Figure] {\label{fig:SNR_adligo} The square root of the
noise spectral density for a {\it signal-recycled} narrowband
Advanced LIGO interferometer with (i) no squeezing
(``Unsqueezed''); (ii) squeezed vacuum injected directly into the
output port of the beamsplitter (``Unfiltered squeezing''); (iii)
squeezed vacuum injected after filtering with an amplitude filter
cavity (``Amplitude filtered squeezing''); and (iv) squeezed
vacuum injected after filtering to get a "Frequency-dependent
squeeze angle".}
\end{figure}

\noindent The sensitivity curve can be shaped by an appropriate
choice of the reflectivity of the signal extraction mirror (SEM)
and the detuning of the signal extraction cavity
(SEC)~\cite{pfspie}. The noise curves shown in
Fig.~\ref{fig:SNR_adligo} are for a narrowband signal-recycled
interferometer where the detuning of the SEC is chosen to place
the optical resonance at 200 Hz. An alternative choice of
detuning, where the carrier light with angular frequency
$\omega_{l}$ obtains a net phase shift of $\frac{\pi}{2}$ in one
pass through the cavity, gives a broadband response. This choice
reduces the variation of the optimal squeeze angle with frequency
and allows for improvement in the performance over a large
frequency range. The broadband interferometer with squeezed input
does not, however, benefit from the amplitude filtering unless the
squeezed state is not pure. We, therefore, limit our discussion to
the narrowband case shown in Fig.~\ref{fig:SNR_adligo}.
Furthermore, we discuss the narrowband signal-recycled
interferometer in term of real frequency, in Hz, since the
normalization to an arm cavity inverse storage time is not
meaningful when the gravtitational-wave signal storage time
depends on both the arm cavity and the (detuned) signal extraction
cavity.

The performance of the squeezed configuration is comparable to the
unsqueezed configuration in the frequency range 10~Hz to 300~Hz,
and considerably better between 300~Hz to 10~kHz.

\section{Filter Performance}
\label{sect:performance}

To evaluate the performance of these amplitude filters with
parameters that can be realized we invoke some simple
astrophysical arguments. The amplitude filters are best suited to
optimizing the gravitational-wave detector performance at high
frequencies without compromising the low-frequency performance as
severely as a fixed squeeze angle input with no filtering. To this
end, three frequency bands are of interest: (i) high frequencies
$\Omega/\gamma \sim 10$ where we expect to carry out searches for
GW emission from pulsars, for example; (ii) the minimum in $S_h$
around $\Omega/\gamma \sim 1$ which is important for detection of
radiation from inspiraling binary systems; and (iii) low
frequencies $\Omega/\gamma < 1$ which are especially important for
detection of a stochastic gravitational-wave background of
cosmological origin by correlating the outputs of two spatially
separated terrestrial detectors~\footnote{We recall that $\gamma =
2 \pi \times 100 {\rm ~Hz}$ for the conventional interferometer
considered here.}. The performance in each of these frequency
bands is characterized by the signal from the targeted source
compared with the noise in the detectors. In Table~\ref{tab:SNR},
we compare the sensitivity or signal-to-noise ratio (SNR) for
pulsars, binary neutron star inspirals and a stochastic background
for several amplitude filter cavity configurations with a
conventional interferometer with varying input power levels.

Most generally, the square of the signal-to-noise ratio (assuming
optimum filtering) is given by~\cite{thorne300}.

\begin{equation}
\left(\frac{S}{N} \right)^2 \propto \int_{0}^{+\infty}
\frac{|h(f)|^2}{S_h(f)}\, df  \label{eqn:SNR}
\end{equation}

\noindent where $h(f)$ is the Fourier transform of the strain
signal from the source and $S_h(f)$ is the single-sided noise
spectral density of the detector.

For \emph{periodic} sources such as a pulsar at a fixed frequency
$\Omega_s$, and ignoring any other modulation effects, the SNR can
be expressed as

\begin{equation}
\left(\frac{S}{N} \right)_{per}^2 \propto \int_{0}^{+\infty}
\frac{\delta(2 \pi \Omega_s)}{S_h(f)}\, df \label{eqn:periodic}
\end{equation}

\noindent where $\delta(2 \pi \Omega_s)$ is a Kronecker delta
function at $\Omega_S$. For our purposes it is convenient to
assume $\Omega_s = 10 \, \gamma$, and the signal-to-noise ratio is
simply proportional to the inverse of the detector noise spectral
density at that frequency.

For the \emph{inspiral} phase of binary neutron star systems, the
Newtonian quadrupole approximation gives $|h(f)|^2 \propto
f^{-7/3}$. We estimate the SNR for inspiraling neutron stars by

\begin{equation}
\left(\frac{S}{N} \right)_{ins}^2 \propto \int_{10 {\rm
~Hz}}^{+\infty} \frac{1}{f^{7/3} \, S_h(f)}\, df
\label{eqn:inspiral}
\end{equation}

\noindent where $S_h(f)$ is given by the curves in
Figures~\ref{fig:SNR} and \ref{fig:SNR_adligo} and a lower cut-off
frequency of $\Omega/\gamma = 0.1$ is used to account for seismic
noise~\footnote{Strictly speaking, there is also an upper cut-off
frequency associated with the innermost stable circular orbit
(ISCO) of the binary system. Above $f_{ISCO} \approx 4400 {\rm
~Hz} \left(M/M_{\odot} \right)^{-1}$, where $M/M_{\odot}$ is the
total mass of the binary system per solar mass, the binary system
enters the merger phase and the spectrum of $|h(f)|^2$ is not
expected to retain a $f^{-7/3}$ dependence. This upper cut-off
frequency does not affect the estimate of the SNR here.}.

For a \emph{stochastic background} of cosmological origin, the
spectrum of gravitational-waves is given by

\begin{equation}
\left |h(f) \right|^2 = \frac{3 H_0^2}{10 \pi^2} f^{-3} E_{gw}(f)
\label{eqn:stochastic}
\end{equation}

\noindent where $H_0$ is the present day Hubble expansion rate and
$E_{gw}(f)$ is the gravitational-wave energy density per
logarithmic frequency, divided by the critical energy density
required to close the Universe. Assuming that $E_{gw}(f) = {\rm
~constant~} = E_0$, and that the waves are isotropic, unpolarized,
stationary and Gaussian, we arrive at $\left |h(f) \right|^2
\propto f^{-3/2}$. Limits on the stochastic background can be set
by cross-correlating the outputs of two detectors. To ensure that
noise in the detectors is not correlated by shared local effects,
we use two widely separated detectors, e.g. detectors at the two
LIGO Observatories, which are nearly co-planar and co-aligned but
separated by a distance $d_s = 3001$~km. The cross-correlation
depends on an additional quantity, $\eta(f)$, the overlap
reduction function, which characterizes the reduction in
sensitivity due to the separation time delay and relative
orientation of the detectors. For the LIGO detectors in Louisiana
and Washington, $\eta(f)$ can be expressed in terms of Bessel
functions, $J_i \,$: $\eta(f) = -0.124842 \, J_0\left(2 \pi f
d_s/c \right) - 2.90014 \, J_1\left(2 \pi f d_s/c \right) +
3.00837 \, J_2\left(2 \pi f d_s/c \right)$ \cite{flanagan}, which
gives $\eta(f=0 {\rm ~Hz~}) \approx 0.9$ with a sharp reduction
above 50~Hz. The square of the signal-to-noise ratio for the
stochastic background obtained by cross-correlating two detectors
with noise spectra, $S_{h {\rm 1}}(f)$ and $S_{h {\rm 2}}(f)$, is
given by

\begin{eqnarray}
\displaystyle \left(\frac{S}{N} \right)_{sb}^2 &\propto&
\displaystyle \int_{-\infty}^{+\infty}
\frac{\eta(f)^2}{f^6 \, S_{h {\rm 1}}(f) \, S_{h {\rm 2}}(f)}\, df \\
&\approx& \displaystyle \int_{-\infty}^{+\infty}
\frac{\eta(f)^2}{f^6 \, S_h^2(f)} \, df \label{eqn:stochasticSNR}
\end{eqnarray}

\noindent The result of Eqn.~(\ref{eqn:stochasticSNR}) assumes
that the noise spectra of the two detectors are Gaussian,
stationary and identical, i.e. $S_{h {\rm 1}}(f) = S_{h {\rm
2}}(f)$, and that optimal filtering was used~\cite{allen}.

\begin{table*}[ht]
\begin{tabular}{lcccc}
\hline\hline \vspace{-0.2cm}\\
Configuration & \hspace{0.3cm} $\displaystyle
\frac{\gamma_{f}}{\gamma}$ & \hspace{0.3cm} Stochastic & \hspace{0.3cm} NS Inspiral & \hspace{0.3cm} Periodic \vspace{0.1cm}\\
\hline\hline \vspace{-0.25cm}\\
Conventional interferometer & \hspace{0.3cm} -- & \hspace{0.3cm} $1.00$ & \hspace{0.3cm} $1.00$ & \hspace{0.3cm} $1.00$\\
Unfiltered fixed-angle squeeze & \hspace{0.3cm} -- & \hspace{0.3cm} $0.32$ & \hspace{0.3cm} $1.16$ & \hspace{0.3cm} $3.16$\\
\hline \vspace{-0.25cm}\\
Single filter & \hspace{0.3cm} $1$ & \hspace{0.3cm} 0.89 & \hspace{0.3cm} 1.00 & \hspace{0.3cm} 3.03\\
Single filter & \hspace{0.3cm} $5$ & \hspace{0.3cm} 0.99 & \hspace{0.3cm} 0.98 & \hspace{0.3cm} 1.89\\
\hline \vspace{-0.25cm}\\
Series filter & \hspace{0.3cm} $1/\sqrt{2}$ & \hspace{0.3cm} 0.98 & \hspace{0.3cm} 1.02 & \hspace{0.3cm} 3.03 \\
Series filter & \hspace{0.3cm} $5/\sqrt{2}$ & \hspace{0.3cm} $1.00$ & \hspace{0.3cm} 1.01 & \hspace{0.3cm} 1.86 \\
\hline \vspace{-0.25cm}\\
Parallel filter & \hspace{0.3cm} $\sqrt{2}$ & \hspace{0.3cm} 0.89 & \hspace{0.3cm} 1.09 & \hspace{0.3cm} 1.12 \\
Parallel filter & \hspace{0.3cm} $5\,\sqrt{2}$ & \hspace{0.3cm} $0.99$ & \hspace{0.3cm} 0.98 & \hspace{0.3cm} 2.24 \\
\hline \vspace{-0.25cm}\\
FD squeeze & \hspace{0.3cm} -- & \hspace{0.3cm} 3.16 & \hspace{0.3cm} 3.16 & \hspace{0.3cm} 3.16 \\
\hline\hline
\end{tabular}
\caption{Comparison of performance of a conventional
interferometer with different filter configurations using three
criteria: the signal-to-noise ratios for detecting (i) a
stochastic background of gravitational-waves, (ii) the inspiral
phase of a neutron star binary system (NS SNR), and (iii) a
periodic source at $\Omega/\gamma=10$ (which is simply the inverse
strain spectral density, $1/\sqrt{S_h}$, at that frequency). We
use the square root of the SNRs defined in
Eqns.~(\ref{eqn:periodic}), (\ref{eqn:inspiral}) and
(\ref{eqn:stochastic}) with $S_h$ corresponding to the various
interferometer and filter configurations listed. All SNRs are
normalized to that of a conventional interferometer with no
squeezing. The series and parallel filter configurations are
described in Section~\ref{sect:multiplefilts}. \label{tab:SNR}}
\end{table*}

\begin{table*}[t]
\begin{tabular}{lcccc}
\hline\hline \vspace{-0.2cm}\\
Configuration & \hspace{0.3cm} $\displaystyle
\frac{\gamma_{f}}{100 {\rm ~Hz}}$ & \hspace{0.2cm} NS Inspiral &
\hspace{0.3cm} $\displaystyle
\frac{1}{\sqrt{S_{h}}}\left(\frac{\Omega}{2 \pi} = 1 {\rm ~kHz}
\right)$ & \hspace{0.3cm} $\displaystyle
\frac{1}{\sqrt{S_{h}}}\left(\frac{\Omega}{2 \pi} = 10 {\rm ~kHz}
\right)$
\vspace{0.1cm}\\
\hline\hline \vspace{-0.25cm}\\
SR interferometer & \hspace{0.3cm} -- & \hspace{0.3cm} $1.00$ & \hspace{0.3cm} $1.00$ & \hspace{0.3cm} $1.00$\\
Unfiltered fixed-angle squeeze & \hspace{0.3cm} -- & \hspace{0.3cm} $0.654$ & \hspace{0.3cm} $3.16$ & \hspace{0.3cm} $3.08$\\
\hline \vspace{-0.25cm}\\
Single filter & \hspace{0.3cm} $3$ & \hspace{0.3cm} $0.854$ & \hspace{0.3cm} $2.39$ & \hspace{0.3cm} $3.07$\\
Single filter & \hspace{0.3cm} $5$ & \hspace{0.3cm} $0.924$ & \hspace{0.3cm} $1.89$ & \hspace{0.3cm} $3.05$\\
Single filter & \hspace{0.3cm} $10$ & \hspace{0.3cm} $0.974$ & \hspace{0.3cm} $1.35$ & \hspace{0.3cm} $2.96$\\
\hline \vspace{-0.25cm}\\
FD squeeze & \hspace{0.3cm} -- & \hspace{0.3cm} 3.16 & \hspace{0.3cm} 3.16 & \hspace{0.3cm} $3.16$\\
\hline\hline
\end{tabular}
\caption{Comparison of performance of a narrowband signal-recycled
interferometer with different filter configurations using two
criteria: The signal-to-noise ratio for (i) inspiraling neutron
star binaries (NS Inspiral), and (ii) periodic sources radiating
at 1 kHz and 10 kHz, respectively. All SNRs are normalized to that
of a signal-recycled interferometer with no squeezing.
\label{tab:SNR_adligo}}
\end{table*}

Table~\ref{tab:SNR} and Fig.~\ref{fig:SNR} show that the best
overall configuration is no doubt the frequency-dependent squeeze
angle. Assuming that the interferometers are operated at the
maximum power possible, then the only way to improve the high
frequency noise is to use squeezing. If no filtering is used, the
binary inspiral SNR is significantly reduced. If an amplitude
filter is used in conjunction with the squeezed input, a moderate
reduction in binary inspiral SNR is traded off against the
benefits of squeezing at higher frequencies. The choice of
bandwidth of the filter influences this trade off. Furthermore,
the amplitude filtered squeezing can be used to increase
sensitivity to binary inspirals by lowering the power, and using
the squeezing to not completely worsen the high frequency noise.
In the likely event that the multiple kilometer-scale filter
cavities needed to achieve the frequency-dependent squeeze angle
are not feasible in the upcoming generation of interferometers,
amplitude filters such as the ones we propose are promising
candidates for broadband improvement in the detector sensitivity.

To explore the feasibility of these filter cavities further, we
give some physical parameters for the amplitude filter cavities.
For a conventional interferometer with initial LIGO parameters
$\gamma \simeq 2 \pi \times 100$~Hz. From Table~\ref{tab:SNR}, we
see that $\gamma \lesssim \gamma_f \lesssim 5\,\gamma$, implying
that the filter cavity linewidth is typically equal to, or a few
times larger, than the arm cavity linewidth, or $500\,{\rm Hz}
\gtrsim \gamma_f/(2 \pi) \gtrsim 100\,{\rm Hz}$. For a 10 meter
long filter cavity, this would correspond to a finesse of 15000 to
75000, or average losses of 70 to 14 parts-per-million per mirror.
If the filter cavities can be made longer (upto $\sim 30$~m is
feasible in the output train of LIGO), the limit on the mirror
losses can be accordingly relaxed.

For completeness we evaluate the performance of a narrowband
signal-recycled interferometer as well. As with the conventional
interferometer, the narrowband signal-recycled configuration of
Fig.~\ref{fig:SNR_adligo} and Table~\ref{tab:SNR_adligo} benefits
at all frequencies from squeezed input with a frequency-dependent
squeeze angle. We intentionally use a narrowband configuration to
highlight the differences between it and the conventional
interferometer. It is evident from Fig.~\ref{fig:SNR_adligo} that
for the narrowband signal-recycled configuration, the amplitude
filter gives significant improvement for detection of neutron star
binary inspirals (the SNR is most sensitive to detector noise in
the minimum between 40 and 400 Hz), but would certainly
deteriorate the detector performance for the stochastic background
(the overlap reduction function strongly favors frequencies below
50 Hz). In Table~\ref{tab:SNR_adligo} it is interesting to note
that the filter cavity can provide substantial benefit at high
frequencies even with considerably higher resonance bandwidths,
e.g. 1~kHz.


\section{Extension to multiple filters}
\label{sect:multiplefilts}

In this section we explore methods to reduce the frequency extent
of the transition region between the reduced noise at high
frequencies (due to squeezing) and the increased noise at low
frequencies, that can be made to approach the noise level of an
interferometer with no squeezing.

\subsection{Series filters}
\begin{figure}[b]
\begin{center}
\includegraphics[width=8cm]{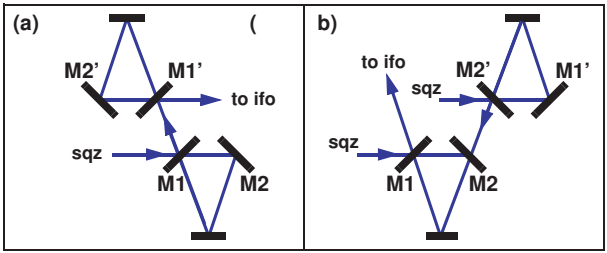}\\
\caption[Figure] {\label{fig:filt2}Panel (a) shows the series
filter, in which the input beam passes through two filters. Panel
(b) shows the parallel filter, in which a filtered beam is
injected instead of vacuum.}
\end{center}
\end{figure}

The most straightforward way to reduce the transition region is to
connect two filters in series, as shown in
Fig.~\ref{fig:filt2}(a). Extending Eqns.~\ref{rhoeq},
\ref{alphaeq} and \ref{betaeq}, we find

\begin{eqnarray}
\alpha_{S} =& |\rho|^{4} &\approx \frac{x^{4}}{x^{4}+2x^{2}+1},\\
\beta_{S} =& 1-|\rho|^{4} &\approx \frac{2x^{2}+1}{x^{4}+2x^{2}+1},\\
x_{1/2} \approx& \left (\sqrt{2}+1\right )^{\frac{1}{2}}.
\end{eqnarray}

\begin{figure}[tbh]
\begin{center}
\includegraphics[width=8cm]{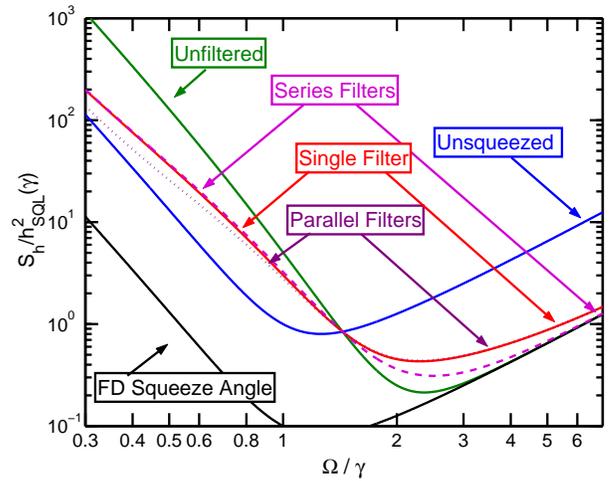}
\caption[Figure] {\label{fig:multSNR} The square root of the noise
spectral density is shown for a conventional interferometer with
(i) no squeezed input ("Unsqueezed"), (ii) Squeezed vacuum
injected ("Unfiltered"); Squeezed input filtered by (iii) a single
filter cavity ("Single Filter"), (iv) series filter cavities
("Series Filter"), (v) parallel filter cavities ("Parallel
Filter"); and (vi) frequency-dependent squeeze angle ("FD Squeeze
Angle").}
\end{center}
\end{figure}

\noindent Comparing the solid curves with the dotted curves in
Fig.~\ref{fig:alphabeta}, we see that the addition of the second
filter has the effect of increasing the attenuation factor, while
shifting the vacuum leakage by a factor of $\sqrt{2}$ in frequency
at frequencies $\Omega \gtrsim \gamma$. By reducing the linewidth
of both filters by a factor $\sqrt{2}$, we obtain nearly the same
vacuum leakage along with an increased attenuation factor, thereby
reducing the size of the transition region. For ease of
comparison, in the "series filter" curve of Fig.~\ref{fig:multSNR}
the linewidth of each filter cavity is reduced by $\sqrt{2}$
compared to an equivalent single filter. We also note that further
gains could be made by double-passing the input squeezed light
through each filter, or, alternatively, reducing the number of
filters required, and hence reducing the complexity.

\subsection{Parallel filters} To decrease the vacuum leakage, we
inject the filtered beam from one filter into the input port of
another filter, as shown in Fig.~\ref{fig:filt2}~(b). We assume
that the input light in each filter is squeezed in the same
quadrature. We also assume that the filters are \emph{lossless}
for simplicity. The output from the combined filters is

\begin{equation}
c_{i} = \rho \left(a_{i} +\tau b_{i}\right) + \tau^{2} v_{i}
\end{equation}

\noindent where $a_{i}$ is the field injected in the first filter,
and $b_{i}$ is the field injected into the second filter. Using
the relation that $|\rho|^{2} + |\tau|^{2} = 1$ for a lossless
filter, we find that

\begin{eqnarray}
\alpha_{P} =& |\rho|^{2}\left(2-|\rho|^{2}\right) &\approx \frac{x^{4}+2x^{2}}{x^{4}+2x^{2}+1},\\
\beta_{P} =& \left(1-|\rho|^{2}\right)^{2} &\approx \frac{1}{x^{4}+2x^{2}+1},\\
x_{1/2} \approx& \left (\sqrt{2}-1\right )^{\frac{1}{2}}.
\end{eqnarray}

\noindent As shown in the dashed curves of
Fig.~\ref{fig:alphabeta}, the parallel filter has the effect of
decreasing the vacuum leakage, while shifting the attenuation
factor by a factor of $\sqrt{2}$ in frequency at frequencies
$\Omega \lesssim \gamma$. By increasing the linewidth of both
filters by a factor $\sqrt{2}$, we can obtain nearly the same
attenuation factor along with reduced vacuum leakage, once again
reducing the size of the transition region.

\subsection{Filter Performance}
It is instructive to evaluate the sensitivity of the
interferometer with squeezed input filtered by the multiple cavity
configurations. For conciseness, we apply multiple filters only to
the conventional interferometer described in Section
\ref{sect:conventional}. The performance measures are listed in
the series and parallel filter sections of Table~\ref{tab:SNR},
where we apply the same criteria as those described in Section
\ref{sect:performance}, namely, the signal-to-noise ratios for
detecting gravitational radiation from (i) a stochastic background
using widely separated detectors, (ii) an inspiraling neutron star
binary system, and (iii) a perfectly periodic source at
$\Omega/\gamma = 10$.

\section{Conclusion}

\label{sect:concl} Recognizing the operational complexity of using
kilometer-scale filter cavities in conjunction with a squeezed
input in long-baseline gravitational-wave interferometers, we have
proposed an alternative type of filter cavity that acts as a
high-pass filter for the squeeze amplitude. We evaluate the
performance of these amplitude filters with parameters that can be
realized and find them to be effective in improving the
high-frequency performance of a squeezed input gravitational-wave
detector without drastically compromising the low-frequency
sensitivity. From Table.~\ref{tab:SNR}, we see that that
significant improvements can be achieved with the
amplitude-filtered squeezed input interferometer compared with an
interferometer with no squeezing, or with a frequency-independent
squeezed-input interferometer, depending on the target
astrophysical source and the interferometer and filter parameters.
The amplitude filters do not give the broadband improvement
afforded by the (multiple) kilometer-scale filter cavities that
give a frequency-dependent squeeze angle, but they are an
attractive alternative since they are a few meters in length and
require finesses well under $10^5$, making them more feasible in
the output train of gravitational-wave detectors. Moreover, the
amplitude filters can suppress noise in excess of the
anti-squeezed quantum-limited noise. We also point out that some
of the broadband benefit afforded by the frequency-dependent
squeeze angle -- or any other filtering scheme -- is likely to be
compromised by other noise sources, e.g., thermal noise, which
were not considered
in this work.\\

\acknowledgments We thank our collaborators, Alessandra Buonanno
and Yanbei Chen, and our colleagues at the LIGO Laboratory,
especially Keisuke Goda and David Ottaway, for stimulating
discussions. We also thank Yanbei Chen for valuable comments on
this manuscript. We gratefully acknowledge support from National
Science Foundation grants PHY-0107417 and PHY-0300345.

\end{document}